\documentclass[12pt]{article}

\usepackage[dvipdfmx]{graphicx}
\usepackage{subcaption}
\usepackage{listings,jvlisting}
\usepackage{amsmath,amssymb}
\usepackage{bm}
\usepackage{graphicx, color}
\usepackage{wrapfig}
\usepackage{cite}
\usepackage{braket}
\usepackage{comment}
\usepackage{hyperref}

\title{
\begin{flushright}
\ \\*[-80pt]
\begin{minipage}{0.2\linewidth}
\normalsize
HUPD-2306 \\*[50pt]
\end{minipage}
\end{flushright}
{\Large \bf
$S_4$ Lepton Flavor Model with 3HDM
\\*[20pt]}}

\author{
\centerline{Yukimura~Izawa $^{1,2}$\footnote{izawa-yukimura@hiroshima-u.ac.jp}}
\\*[20pt]
\centerline{
\begin{minipage}{\linewidth}
\begin{center}
$^1${\it \normalsize
Physics Program, Graduate School of Advanced Science \\ and Engineering,~Hiroshima~University, \\
Higashi-Hiroshima~739-8526,~Japan \\*[5pt]}
$^2${\it \normalsize
International Institute for Sustainability
with Knotted Chiral Meta Matter/WPI-SKCM2, Hiroshima University, \\
2-313 Kagamiyama, Higashi-Hiroshima City, Hiroshima, Japan 739-0046}
\end{center}
\end{minipage}
}
\\*[50pt]
}

\date{
\centerline{\small \bf Abstract}
\begin{minipage}{0.9\linewidth}
\medskip
\medskip
\small
We prsent a lepton flavor model with $S_4$ and $U(1)_{\mathrm{FN}}$ symmetry.
The left-handed leptons are assigned as a triplet under the $S_4$ symmetry, the right-handed electron and muon as $S_4$ doublets, and the right-handed tauon as an $S_4$ singlet.
We introduce three right-handed Majorana neutrinos that are charged under the $S_4$ symmetry and two scalar fields.
Additionally, the three Higgs doublets are assigned as an $S_4$ triplet, and we analyze the scalar potential to get the conditions for the Higgs VEVs.
We numerically calculate the PMNS matrix, and give strong predictions for the mixing angle $0.486<\sin\theta_{23}<0.603$ and the Dirac CP phase $60.12^\circ<|\delta_{CP}|<76.47^\circ$.
Our predictions for the lightest neutrino mass $m_{light}\simeq5.53~[\mathrm{meV}]$ and the effective Majorana neutrino mass $m_{ee}\simeq6.33~[\mathrm{meV}]$ from the neutrinoless
double beta decay experiments are relatively close to upper limits.
Lastly, we obtain the sum of neutrino masses $m_1+m_2+m_3\simeq66.1~[\mathrm{meV}]$, and two Majorana phases $\eta_1$ and $\eta_2$.
\end{minipage}
}
\begin{document}
\begin{titlepage}
	\maketitle
\thispagestyle{empty}
\end{titlepage}
\newpage

%------------------------------------------------------------------------------%
%--------------------------------   Introduction   --------------------------------%
%------------------------------------------------------------------------------%
\section{Introduction}
The Standard Model (SM) of elementary particles successfully describes the known particles and their interactions.
It has explained the mechanism of electroweak symmetry breaking and the mechanism of mass acquisition for gauge bosons and fermions.
With the discovery of the Higgs particle in 2012~\cite{ATLAS:2012yve, CMS:2012qbp}, the SM is believed to be a correct theory.
However, there are some mysteries in the SM, for example the generational structure of fermions.
% The SM fermions, quarks and leptons, have a generation structure.
That structure comes from the magnitude of the Yukawa interactions, which describe the interaction between charged fermions and the Higgs boson in the SM.
The Yukawa interactions are simply parameterized to match experimental data.
Therefore, it is not possible to theoretically explain the difference in masses and mixing of flavors between generations in the SM.
In particular, the flavor mixing of leptons is predicted to be larger than that of quarks and the magnitude of the CP phase of the lepton sector is not yet precisely known.

Leptons have another mystery related to neutrinos.
In the SM the neutrinos have zero mass, because there are no right-handed neutrinos.
However, the discovery of neutrino oscillations has shown that neutrinos have tiny masses.
Although neutrino masses can not be explained by the SM, the neutrino mass matrix is very important when considering lepton flavor mixing.
In addition, it is not yet known whether neutrinos are Dirac or Majorana particles.
If the neutrinos are Majorana particles, then two additional CP phases called Majorana phases are added, the magnitude of which is not yet known.

Therefore, many studies have been conducted to understand the generation structure.
For example, the Froggatt-Nielsen (FN) mechanism introduced global a $U(1)_\mathrm{FN}$ symmetry~\cite{Froggatt:1978nt} to provide a natural explanation for the fermion mass hierarchies.
% Because of the $U(1)_\mathrm{FN}$ symmetry, it provides a natural explanation for the fermion mass hierarchies.
Also, several studies of non-Abelian discrete symmetries (see for review~\cite{Ishimori:2010au, Ishimori:2012zz, Kobayashi:2022moq}) have been used to explain flavor mixing~\cite{Ma:2001dn, Altarelli:2005yp, Altarelli:2005yx,Brahmachari:2008fn, Altarelli:2010gt, Ishimori:2010fs, King:2013eh, King:2014nza}.
The Yukawa couplings are controlled by these symmetries, and it leads to a natural explanation for the mixing angles of lepton.
%One of the successful flavor models was proposed by G.~Altarelli and F.~Feruglio~\cite{Altarelli:2005yp,Altarelli:2005yx}.
%The Altarelli and Feruglio (AF) model show the TBM by using the non-Abelian discrete symmetry $A_4$.

Much work has also been done to explain neutrino masses~\cite{Minkowski:1977sc, Yanagida:1979as, Gell-Mann:1979vob, Mohapatra:1979ia, Schechter:1980gr, Yanagida:1980xy}.
The masses of the lighter left-handed neutrinos are often explained using heavy right-handed Majorana neutrinos and the seesaw mechanism.
Majorana neutrinos are being searched for using neutrinoless double beta ($0\nu\beta\beta$) decay at experiments like KamLAND-Zen~\cite{KamLAND-Zen:2016pfg,KamLAND-Zen:2022tow}.
These experiments have provided an upper limit to the effective Majorana neutrino mass.
According to this paper~\cite{KamLAND-Zen:2022tow}, the inverted ordering region is being explored.

In building our flavor model, in addition to the non-Abelian discrete and the $U(1)_\mathrm{FN}$ symmetries, we focus on the Higgs sector.
Many multi-Higgs models have appeared in recent times.
It is natural to assume that not only fermions but also Higgs sector has a similar generation structure, and there are many previous studies of that type~\cite{Lavoura:2007dw, deAdelhartToorop:2010jxh, Ivanov:2012ry, Degee:2012sk, GonzalezFelipe:2013xok, GonzalezFelipe:2013yhh, Ivanov:2012fp, Keus:2013hya, Ivanov:2014doa,Ivanov:2017dad, Das:2018qyt, Das:2019ntw, Ivanov:2020jra, Buskin:2021eig, Carrolo:2022oyg, Vergeest:2022mqm, CarcamoHernandez:2022vjk, Izawa:2022viu}.
We consider a three Higgs doublets model(3HDM)~\cite{Weinberg:1976hu}, which means the number of Higgs doublets is three and has a flavor symmetry.
% which is the same number of fermion generations.
% We construct a flavor model of leptons in which the Higgs sector also has the flavor symmetry.
This means we assume the Higgs sector follows the same generation structure as the fermions.
In a previous study, we considered an $A_4$ symmetry, which is one possible non-Abelian discrete symmetry, and the 3HDM to construct a flavor model~\cite{Izawa:2022viu}.
In that study, we could reproduce the lepton mixing angles and give predictions for the CP violating phase and the lightest neutrino mass.
However, we only reproduced the neutrino masses in the inverted hierarchy.
As mentioned above, the inverted hierarchy is currently being explored and unfavor~\cite{KamLAND-Zen:2022tow}.

In this paper, we consider the $S_4$ symmetry that is also a possible non-Abelian discrete symmetry.
The advantage of the $S_4$ symmetry is that the $S_4$ symmetry contains one doublet that is not included in the $A_4$ symmetry.
Because of that we can build an $S_4$ lepton flavor model with $U(1)_\mathrm{FN}$ and 3HDM, which reproduce neutrino masses in the normal hierarchy. 
Additionally, our model predicts the mixing angles and the magnitude of Dirac CP phase of leptons.
We also predict the lightest neutrino mass, the sum of neutrino masses, the effective Majorana neutrino mass at the $0\nu\beta\beta$ decay experiment and the two Majorana phases in normal-ordering.
%The T2K and NO$\nu $A experiments have confirmed the neutrino oscillation in the $\nu _\mu \to \nu _e$ appearance events~\cite{T2K:2013bqz,T2K:2013ppw,NOvA:2016kwd},
%\cite{T2K:2013bqz}--\cite{NOvA:2016kwd}, which are one of the clues of the new physics beyond the SM such as the Dirac CP violating phase for the lepton sector by combining the data of the reactor neutrino  experiments~\cite{DayaBay:2012fng,RENO:2012mkc}.
%The KamLAND-Zen~\cite{KamLAND-Zen:2016pfg,KamLAND-Zen:2022tow}, GERDA\cite{GERDA:2013vls,GERDA:2018pmc}, and CUORE~\cite{CUORE:2017tlq,CUORE:2019yfd} experiments also provide us the significant informations which are whether the neutrinos are Dirac or Majorana particles, the lepton number violation, and Majorana phases if the neutrinos are Majorana particles.
%Thus the neutrino oscillation experiments go into a new phase of the precise determinations of the lepton mixing angles, the neutrino mass squared differences, and the CP violating phases.

The rest of this paper is organized as follows.
In Section~\ref{sec:model}, we present the lepton flavor model and obtain mass matrices.
In Section~\ref{sec:Numerical}, we show the numerical analysis of our flavor model.
In Section~\ref{sec:poten_analysis}, we show the scalar potential analysis.
In Section~\ref{sec:Summary}, we summarize this paper.
We give the brief introduction of the $S_4$ symmetry and show the multiplication rule of the $S_4$ group in Appendix~\ref{sec:intro_S4}.
We shortly introduce the 3HDM with $S_4$ symmetry in Appendix~\ref{sec:3HDM_S4}.

%-----------------------------------------------------%
%------------------model------------------------------%
%-----------------------------------------------------%
\section{Lepton flavor model with $S_4$ symmetry} \label{sec:model}
In this section, we present a lepton flavor model with $S_4$ symmetry, and we show the mass matrices of charged leptons and neutrinos.

The $S_4$ symmetry is the symmetry of the $S_4$ group, which is the symmetric group of order 4.
The $S_4$ group has two types of singlets $\mathbf{1}$, $\mathbf{1'}$, one type of doublet $\mathbf{2}$, and two types of triplets $\mathbf{3}$, $\mathbf{3'}$, as noted in Appendix~\ref{sec:intro_S4}.
We assign the left-handed leptons as an $S_4$ triplet, the right-handed electron and muon as $S_4$ doublets, and the right-handed tauon as an $S_4$ singlet.
Then, we suppose three right-handed Majorana neutrinos to obtain neutrino masses with the seesaw mechanism.
Uniquely, the right-handed electron neutrino is assigned as an $S_4$ singlet, and the right-handed muon neutrino and tauon neutrinos are assigned as $S_4$ doublets.

Here, we introduce the three Higgs doublets that are assigned as an $S_4$ triplet.
This means the Higgs sector has a similar generation structure to fermions.
Additionally, we introduce two scalar fields $\Theta$ and $X$, where the field $\Theta$ is an $S_4$ singlet and the field $X$ is an $S_4$ doublet.
This allows us to preform a precise reproduction of experimental results.
We impose the $U(1)_{\mathrm{FN}}$ symmetry on the added scalar fields and right-handed lepton doublet, and we limit the couplings of scalar fields.
In Table 1, we summarize the particle assignments of the $SU(2)_L$, $S_4$ and $U(1)_{\mathrm{FN}}$ symmetries.
\begin{table}[h]
  \centering
  \begin{tabular}{|c||c|c|c|c|c|c|c|c|}
    \hline 
    \rule[14pt]{0pt}{0pt}
    & $\bar{\ell}=(\bar{\ell}_e,\bar{\ell}_\mu,\bar{\ell}_\tau)$ & $\ell_R=(e_R,\mu_R)$ &$\tau_R$ & $\nu_{eR}$ & $\nu_R=(\nu_{\mu R},\nu_{\tau R})$ & $\phi=(\phi_1,\phi_2,\phi_3)$ & $\Theta$ & $X$ \\
    \hline 
    \rule[14pt]{0pt}{0pt}
    $SU(2)_L$ & $2$ & $1$ & $1$ & $1$ & $1$ & $2$ & $1$ & $1$ \\
    $S_4$ & ${\bf 3}$ & ${\bf 2}$ & ${\bf 1}$ & ${\bf 1}$ & ${\bf 2}$ & ${\bf 3}$ & ${\bf 1}$ & ${\bf 2}$\\
    $U(1)_{\mathrm{FN}}$ & $0$ & $+1$ & $0$ & $0$ & $0$ & $0$ & $-1$ & $-1$ \\
    \hline
  \end{tabular}
  \caption{The charge assignments of $SU(2)_L\times S_4 \times U(1)_{\mathrm{FN}} $ symmetry in our model.}
%, $\ell$ is 3 lepton doublets.}
  \label{tab:model}
\end{table}

We can write down the Lagrangian for Yukawa interactions and Majorana mass term in our model.
The $SU(2)_L\times S_4\times U(1)_{\mathrm{FN}}$ invariant Lagrangian is,
%Yukawa interaction and Majorana mass term in Lagrangian for this model is Table~\ref{tab:model}
\begin{equation}\label{eq:lagrangian_all}
  \mathcal{L}_Y = \mathcal{L}_{\ell} + \mathcal{L}_D + \mathcal{L}_M + h.c. ,
\end{equation}
where,
\begin{align}
	\mathcal{L}_{\ell} &= \frac{y_{e\mu}}{\Lambda}\bar{\ell}\phi\ell_R\Theta+ y_\tau\bar{\ell}\phi\tau_R 
+ \frac{y_\ell}{\Lambda}\bar{\ell}\phi\ell_R X,\label{eq:lagrangian_charged} \\
	\mathcal{L}_D &= y_{De}\bar{\ell}\tilde{\phi}\nu_{eR}+y_{D\mu\tau}\bar{\ell}\tilde{\phi}\nu_{R},\label{eq:lagrangian_Dirac} \\
	\mathcal{L}_M &= \frac{1}{2}M_{eR}\bar{\nu}_{eR}^C\nu_{eR}+\frac{1}{2}M_{\mu\tau R}\bar{\nu}_R^C\nu_{R}. \label{eq:lagrangian_Majorana}
\end{align}
Note that $y_{e\mu} , y_\tau , y_\ell$, $y_{De}$ and $y_{D\mu\tau}$ are Yukawa couplings and, $M_{eR}$ and $M_{\mu\tau R}$ are the right-handed Majorana neutrino masses.
We consider the Yukawa coupling $y_{e\mu} , y_\tau, y_\ell$ as real numbers, and $y_{De}$ and $y_{D\mu\tau}$ to be complex numbers.
At the high-energy scale, we take the real VEVs of $\Theta$ and $X$ as $\langle \Theta \rangle = \Theta_0,  \langle X \rangle =(\mathrm{X}_1,0)$.
Then, after spontaneous symmetry breaking (SSB), the three Higgs doublets have the real VEVs $\langle \phi_i \rangle = \begin{pmatrix} 0 \\ \frac{1}{\sqrt{2}}v_i \end{pmatrix} $ $(i=1,2,3)$.
For the charged lepton sector Eq.~\eqref{eq:lagrangian_charged}, we derive the Yukawa interactions following the $S_4$ multiplication rules Eq.~\eqref{eq:multi_rule_rep}-\eqref{eq:multi_rule_3*3} in Appendix~\ref{sec:intro_S4};
\begin{align}
	\frac{y_{e\mu}}{\Lambda}\bar{\ell}\phi\ell_R\Theta
	&= \frac{y_{e\mu} \Theta}{\Lambda} \begin{pmatrix} \bar{\ell}_e \\ \bar{\ell}_\mu \\ \bar{\ell}_\tau \end{pmatrix} \times \begin{pmatrix} \phi_1 \\ \phi_2 \\ \phi_3 \end{pmatrix} \times \begin{pmatrix} e_R \\ \mu_R  \end{pmatrix} \notag \\
	&= \frac{y_{e\mu} \Theta}{\Lambda}\Big[ \frac{1}{\sqrt{2}}(\bar{\ell}_\mu \phi_2 -\bar{\ell}_\tau \phi_3)e_R+\frac{1}{\sqrt{6}}(-2\bar{\ell}_e \phi_1+\bar{\ell}_\mu \phi_2+\bar{\ell}_\tau \phi_3)\mu_R\Big] \notag \\
	&\rightarrow \frac{y_{e\mu}\Theta_0}{\Lambda} \Big[ \frac{1}{2}(\bar{\mu}_L v_2 -\bar{\tau}_Lv_3)e_R+\frac{1}{2\sqrt{3}}(-2\bar{e}_Lv_1+\bar{\mu}_Lv_2+\bar{\tau}_Lv_3)\mu_R\Big], \label{eq:cal_lagrangian_charged_emu} \\
	y_\tau\bar{\ell}\phi\tau_R
	&=y_\tau \begin{pmatrix} \bar{\ell}_e \\ \bar{\ell}_\mu \\ \bar{\ell}_\tau \end{pmatrix} \times \begin{pmatrix} \phi_1 \\ \phi_2 \\ \phi_3 \end{pmatrix} \tau_R \notag \\
	&=y_\tau (\bar{\ell}_e \phi_1+\bar{\ell}_\mu \phi_2+\bar{\ell}_\tau \phi_3)\tau_R \notag \\
	&\rightarrow \frac{y_\tau}{\sqrt{2}}(\bar{e}_Lv_1+\bar{\mu}_Lv_2+\bar{\tau}_Lv_3)\tau_R, \label{eq:cal_lagrangian_charged_tau} \\
	\frac{y_\ell}{\Lambda}\bar{\ell}\phi\ell_R X
	&= \frac{y_\ell}{\Lambda} \begin{pmatrix} \bar{\ell}_e \\ \bar{\ell}_\mu \\ \bar{\ell}_\tau \end{pmatrix} \times \begin{pmatrix} \phi_1 \\ \phi_2 \\ \phi_3 \end{pmatrix} \times \begin{pmatrix} e_R \\ \mu_R  \end{pmatrix} \times \begin{pmatrix} X_1 \\ X_2  \end{pmatrix} \notag \\
	&= \frac{y_{\ell1}}{\Lambda}(\bar{\ell}_e\phi_1+\bar{\ell}_\mu\phi_2+\bar{\ell}_\tau\phi_3)(e_R X_1+\mu_R X_2) \notag \\
	&\quad \, +\frac{y_{\ell2}}{\Lambda}\Big[\frac{1}{\sqrt{2}}(\bar{\ell}_\mu\phi_2-\bar{\ell}_\tau\phi_3)(e_R X_2+\mu_R X_1)+\frac{1}{\sqrt{6}}(-2\bar{\ell}_e\phi_1+\bar{\ell}_\mu\phi_2+\bar{\ell}_\tau\phi_3)(e_R X_1-\mu_R X_2)\Big] \notag \\
	&\rightarrow \frac{y_{\ell1}}{\sqrt{2}\Lambda}(\bar{e}_L v_1+\bar{\mu}_L v_2+\bar{\tau}_L v_3)e_R \mathrm{X}_1 \notag \\
	& \quad \, +\frac{y_{\ell2}}{\Lambda}\Big[\frac{1}{2}(\bar{\mu}_L v_2-\bar{\tau}_L v_3)\mu_R \mathrm{X}_1+\frac{1}{2\sqrt{3}}(-2\bar{e}_L v_1+\bar{\mu}_L v_2+\bar{\tau}_L v_3)e_R \mathrm{X}_1 \Big]. \label{eq:cal_lagrangian_charged_chi}
\end{align}
The mass matrix from Eq.~\eqref{eq:cal_lagrangian_charged_emu} and Eq.~\eqref{eq:cal_lagrangian_charged_tau} is $M_{\ell 1}$, where
\begin{align}
	M_{\ell1}=
	\begin{pmatrix}
		0 & -\frac{1}{\sqrt{3}} \frac{y_{e\mu}\Theta_0}{\Lambda} v_1 & \frac{1}{\sqrt{2}}y_\tau v_1 \\
 		\frac{1}{2} \frac{y_{e\mu}\Theta_0}{\Lambda} v_2 & \frac{1}{2\sqrt{3}} \frac{y_{e\mu}\Theta_0}{\Lambda} v_2 & \frac{1}{\sqrt{2}}y_\tau v_2 \\
		 -\frac{1}{2} \frac{y_{e\mu}\Theta_0}{\Lambda} v_3& \frac{1}{2\sqrt{3}} \frac{y_{e\mu}\Theta_0}{\Lambda} v_3 & \frac{1}{\sqrt{2}}y_\tau v_3
	\end{pmatrix}_{LR \, .}
\end{align}
The mass matrix from Eq.~\eqref{eq:cal_lagrangian_charged_chi} is $M_{\ell 2}$, where
\begin{align}
	M_{\ell2}=
	\begin{pmatrix}
		(\frac{1}{\sqrt{2}}y_{\ell1}-\frac{1}{\sqrt{3}}y_{\ell2})v_1 \frac{\mathrm{X}_1}{\Lambda} & 0 & 0 \\
 		(\frac{1}{\sqrt{2}}y_{\ell1}+\frac{1}{2\sqrt{3}}y_{\ell2})v_2 \frac{\mathrm{X}_1}{\Lambda} & \frac{1}{2} y_{\ell2} v_2 \frac{\mathrm{X}_1}{\Lambda} & 0 \\
		(\frac{1}{\sqrt{2}}y_{\ell1}+\frac{1}{2\sqrt{3}} y_{\ell2} )v_3 \frac{\mathrm{X}_1}{\Lambda} & -\frac{1}{2} y_{\ell2} v_3 \frac{\mathrm{X}_1}{\Lambda} & 0 \\
	\end{pmatrix}_{LR \, .}
\end{align}
Then, we obtain the charged lepton mass matrix $M_{\ell }$ as
\begin{align}
	M_{\ell} &= M_{\ell 1} + M_{\ell 2} \notag \\
	&=
	\begin{pmatrix}
		(\frac{1}{\sqrt{2}}y_{\ell1}-\frac{1}{\sqrt{3}}y_{\ell2})v_1 \frac{\mathrm{X}_1}{\Lambda} & -\frac{1}{\sqrt{3}} \frac{y_{e\mu}\Theta_0}{\Lambda} v_1 &\frac{1}{\sqrt{2}} y_\tau v_1 \\
 		\frac{1}{2} \frac{y_{e\mu}\Theta_0}{\Lambda} v_2 + (\frac{1}{\sqrt{2}}y_{\ell1}+\frac{1}{2\sqrt{3}}y_{\ell2})v_2 \frac{\mathrm{X}_1}{\Lambda} & \frac{1}{2\sqrt{3}} \frac{y_{e\mu}\Theta_0}{\Lambda} v_2 +\frac{1}{2} y_{\ell2} v_2 \frac{\mathrm{X}_1}{\Lambda} & \frac{1}{\sqrt{2}}y_\tau v_2 \\
		 -\frac{1}{2} \frac{y_{e\mu}\Theta_0}{\Lambda} v_3 + (\frac{1}{\sqrt{2}}y_{\ell1}+\frac{1}{2\sqrt{3}} y_{\ell2} )v_3 \frac{\mathrm{X}_1}{\Lambda} & \frac{1}{2\sqrt{3}} \frac{y_{e\mu}\Theta_0}{\Lambda} v_3 -\frac{1}{2} y_{\ell2} v_3 \frac{\mathrm{X}_1}{\Lambda} & \frac{1}{\sqrt{2}}y_\tau v_3
 	\end{pmatrix}_{LR \, .} \label{eq:massmat_charged}
\end{align}

Next, we calculate the Dirac neutrino mass matrix from the Lagrangian for Dirac neutrino Yukawa interactions Eq.~\eqref{eq:lagrangian_Dirac}.
As with the charged lepton mass matrix, the Dirac neutrino mass matrix $M_D$ is obtained via the following,
\begin{align}
	M_D=
	\begin{pmatrix}
		\frac{1}{\sqrt{2}}y_{De}v_1 & 0 & -\frac{1}{\sqrt{3}}y_{D\mu\tau}v_1 \\
 		\frac{1}{\sqrt{2}} y_{De}v_2 & \frac{1}{2}y_{D\mu\tau}v_2 & \frac{1}{2\sqrt{3}}y_{D\mu\tau}v_2 \\
		 \frac{1}{\sqrt{2}} y_{De}v_3& -\frac{1}{2}y_{D\mu\tau}v_3 & \frac{1}{2\sqrt{3}}y_{D\mu\tau}v_3
	\end{pmatrix}_.
\end{align}
% We look the right-handed Majorana neutrino mass matrix.
We also derive the right-handed Majorana neutrino mass matrix $M_R$ from Eq.~\eqref{eq:lagrangian_Majorana},
\begin{align}
	M_R=
	\begin{pmatrix}
		M_{eR} & 0 & 0 \\
 		0 & M_{\mu\tau R} & 0 \\
		0 & 0 & M_{\mu\tau R}
	\end{pmatrix}_.
\end{align}
After these calculations of neutrino mass matrices, we use the type-I seesaw mechanism and obtain the left-handed Majorana neutrino mass matrix $m_\nu$:
\begin{align}
	&m_\nu = -M_D M_R^-1 M_D^T  \label{eq:massmat_mnu}, \\
	&(m_\nu)_{ij}=-\frac{e^{2i \phi_{De}}|y_{De}|^2v_i^2}{2M_{eR}}-\frac{e^{2i \phi_{D\mu\tau}}|y_{D\mu\tau}|^2v_i^2}{3M_{\mu\tau R}}, \qquad \,  \,  i=j \,  (i,j=1,2,3), \notag \\
	&(m_\nu)_{ij}=-\frac{e^{2i\phi_{De}}|y_{De}|^2v_iv_j}{2M_{eR}}+\frac{e^{2i \phi_{D\mu\tau}}|y_{D\mu\tau}|^2v_iv_j}{6M_{\mu\tau R}}, \quad i\neq j \, (i,j=1,2,3), \notag
\end{align}
\begin{comment}
\begin{align}
	\setbox0=\vbox{\parindent=-20pt$\displaystyle
	\footnotesize
	=\begin{pmatrix}
		-\frac{e^{2i \phi_{De}}|y_{De}|^2v_1^2}{2M_{eR}}-\frac{e^{2i \phi_{D\mu\tau}}|y_{D\mu\tau}|^2v_1^2}{3M_{\mu\tau R}} & -\frac{e^{2i\phi_{De}}|y_{De}|^2v_1v_2}{2M_{eR}}+\frac{e^{2i \phi_{D\mu\tau}}|y_{D\mu\tau}|^2v_1v_2}{6M_{\mu\tau R}} & -\frac{e^{2i\phi_{De}}|y_{De}|^2v_3v_1}{2M_{eR}}+\frac{e^{2i \phi_{D\mu\tau}}|y_{D\mu\tau}|^2v_3v_1}{6M_{\mu\tau R}} \\
 		-\frac{e^{2i\phi_{De}}|y_{De}|^2v_1v_2}{2M_{eR}}+\frac{e^{2i \phi_{D\mu\tau}}|y_{D\mu\tau}|^2v_1v_2}{6M_{\mu\tau R}} & -\frac{e^{2i\phi_{De}}|y_{De}|^2v_2^2}{2M_{eR}}-\frac{e^{2i \phi_{D\mu\tau}}|y_{D\mu\tau}|^2v_2^2}{3M_{\mu\tau R}} & -\frac{e^{2i\phi_{De}}|y_{De}|^2v_2v_3}{2M_{eR}}+\frac{e^{2i \phi_{D\mu\tau}}|y_{D\mu\tau}|^2v_2v_3}{6M_{\mu\tau R}} \\
		-\frac{e^{2i\phi_{De}}|y_{De}|^2v_3v_1}{2M_{eR}}+\frac{e^{2i \phi_{D\mu\tau}}|y_{D\mu\tau}|^2v_3v_1}{6M_{\mu\tau R}}  & -\frac{e^{2i\phi_{De}}|y_{De}|^2v_2v_3}{2M_{eR}}+\frac{e^{2i \phi_{D\mu\tau}}|y_{D\mu\tau}|^2v_2v_3}{6M_{\mu\tau R}} & -\frac{e^{2i\phi_{De}}|y_{De}|^2v_3^2}{2M_{eR}}-\frac{e^{2i \phi_{D\mu\tau}}|y_{D\mu\tau}|^2v_3^2}{3M_{\mu\tau R}} \notag 
	\end{pmatrix}_,$} 
	\normalsize
	\noindent\scalebox{.9}[1]{\box0}
	\\ \label{eq:massmat_mnu}
\end{align}
\end{comment}
where, $\phi_{De}$ and $\phi_{D\mu\tau}$ are the complex phases from the Yukawa couplings $y_{De}$ and $y_{D\mu\tau}$.

%--------------------------------------------------------------------%
\section{Numerical analysis}
\label{sec:Numerical}
%--------------------------------------------------------------------%
In this section, we numerically obtain the PMNS matrix from the leptons mass matrices.
We show the numerical results such as the mixing angles of leptons, Dirac CP phase, the lightest neutrino mass, the sum of neutrino masses, the effective Majorana neutrino mass at the $0\nu\beta\beta$ decay experiment, and two Majorana phases in normal-ordering.

First, we get the PMNS matrix representing the lepton flavor mixing from the mass matrices of charged leptons and neutrinos obtained in the section \ref{sec:model}. 
We find the unitary matrix $V_\ell$ that diagonalizes the charged lepton mass matrix.
The charged lepton mass matrix $M_{\ell}$ in Eq.~\eqref{eq:massmat_charged} becomes
\begin{align}
	M_{\ell} =
	\begin{pmatrix}
		(\frac{1}{\sqrt{2}}y_{\ell1}-\frac{1}{\sqrt{3}}y_{\ell2})v_1 \mathrm{X'}_1 & -\frac{1}{\sqrt{3}} y'_{e\mu} v_1 &\frac{1}{\sqrt{2}} y_\tau v_1 \\
 		\frac{1}{2} y'_{e\mu} v_2 + (\frac{1}{\sqrt{2}}y_{\ell1}+\frac{1}{2\sqrt{3}}y_{\ell2})v_2 \mathrm{X'}_1 & \frac{1}{2\sqrt{3}} y'_{e\mu} v_2 +\frac{1}{2} y_{\ell2} v_2 \mathrm{X'}_1 & \frac{1}{\sqrt{2}}y_\tau v_2 \\
		 -\frac{1}{2} y'_{e\mu} v_3 + (\frac{1}{\sqrt{2}}y_{\ell1}+\frac{1}{2\sqrt{3}} y_{\ell2} )v_3 \mathrm{X'}_1 & \frac{1}{2\sqrt{3}} y'_{e\mu} v_3 -\frac{1}{2} y_{\ell2} v_3 \mathrm{X'}_1 & \frac{1}{\sqrt{2}}y_\tau v_3
 	\end{pmatrix}_{LR ,} \label{eq:massmat_charged_revise}
\end{align}
where, $\mathrm{X'}_1 \equiv \frac{\mathrm{X}_1}{\Lambda}$, and $y'_{e\mu} \equiv \frac{y_{e\mu}\Theta_0}{\Lambda}$.
We use $M_{\ell}M_{\ell}^\dagger$ to find the unitary matrix $V_\ell$ that diagonalizes $M_{\ell}M_{\ell}^\dagger$.
Similarly, we evaluate the neutrino sector in the same way.
The neutrino mass matrix $m_\nu$ in Eq.~\eqref{eq:massmat_mnu} is rewritten to be
\begin{align}
	&(m_\nu)_{ij}=-\frac{1}{2}e^{2i \phi_{De}}|y'_{De}|^2v_i^2-\frac{1}{3}e^{2i \phi_{D\mu\tau}}|y'_{D\mu\tau}|^2v_i^2, \qquad \,  \,  i=j \,  (i,j=1,2,3), \notag \\
	&(m_\nu)_{ij}= -\frac{1}{2}e^{2i\phi_{De}}|y'_{De}|^2v_iv_j+\frac{1}{6}e^{2i \phi_{D\mu\tau}}|y'_{D\mu\tau}|^2v_iv_j, \quad  i\neq j \, (i,j=1,2,3), \notag
\end{align}
\begin{comment}
\begin{align}
	&\setbox0=\vbox{\parindent=-65pt$\displaystyle
	\footnotesize
	m_\nu=\begin{pmatrix}
		-\frac{1}{2}e^{2i \phi_{De}}|y'_{De}|^2v_1^2-\frac{1}{3}e^{2i \phi_{D\mu\tau}}|y'_{D\mu\tau}|^2v_1^2 & -\frac{1}{2}e^{2i\phi_{De}}|y'_{De}|^2v_1v_2+\frac{1}{6}e^{2i \phi_{D\mu\tau}}|y'_{D\mu\tau}|^2v_1v_2 & -\frac{1}{2}e^{2i\phi_{De}}|y'_{De}|^2v_3v_1+\frac{1}{6}e^{2i \phi_{D\mu\tau}}|y'_{D\mu\tau}|^2v_3v_1 \\
 		-\frac{1}{2}e^{2i\phi_{De}}|y'_{De}|^2v_1v_2+\frac{1}{6}e^{2i \phi_{D\mu\tau}}|y'_{D\mu\tau}|^2v_1v_2 & -\frac{1}{2}e^{2i\phi_{De}}|y'_{De}|^2v_2^2-\frac{1}{3}e^{2i \phi_{D\mu\tau}}|y'_{D\mu\tau}|^2v_2^2 & -\frac{1}{2}e^{2i\phi_{De}}|y'_{De}|^2v_2v_3+\frac{1}{6}e^{2i \phi_{D\mu\tau}}|y'_{D\mu\tau}|^2v_2v_3 \\
		-\frac{1}{2}e^{2i\phi_{De}}|y'_{De}|^2v_3v_1+\frac{1}{6}e^{2i \phi_{D\mu\tau}}|y'_{D\mu\tau}|^2v_3v_1  & -\frac{1}{2}e^{2i\phi_{De}}|y'_{De}|^2v_2v_3+\frac{1}{6}e^{2i \phi_{D\mu\tau}}|y'_{D\mu\tau}|^2v_2v_3 & -\frac{1}{2}e^{2i\phi_{De}}|y'_{De}|^2v_3^2-\frac{1}{3}e^{2i \phi_{D\mu\tau}}|y'_{D\mu\tau}|^2v_3^2 \notag 
	\end{pmatrix}_,$} 
	\normalsize
	\noindent\scalebox{.77}[1]{\box0}
	\\ \label{eq:massmat_mnu_revise}
\end{align}
\end{comment}
where, $|y'_{De}|^2 \equiv \frac{|y_{De}|^2}{M_{eR}}$, and $|y'_{D\mu\tau}|^2 \equiv \frac{|y_{D\mu\tau}|^2}{M_{\mu \tau R}}$.
We use $m_\nu m_\nu^\dagger$ to find the unitary matrix $V_\nu$ that diagonalizes $m_\nu m_\nu^\dagger$.

Next, we perform the numerical analysis with the physical input parameters the masses of the charged leptons $m_e, m_\mu, m_\tau$, and the neutrino mass-squared differences $\Delta m^2_{21}, \Delta m^2_{31}$.
We assume normal hierarchy for the neutrinos and take, from NuFIT 5.2~\cite{Esteban:2020cvm,NuFIT5.2}, the neutrino mass-squared differences $\Delta m^2_{21} = 7.41 \times 10^{-5}~\mathrm{eV^2}$, $\Delta m^2_{31} = 2.507 \times 10^{-3}~\mathrm{eV^2}$ in Table~\ref{tab:NuFIT_data}.
Additionally, we use the PDG data~\cite{ParticleDataGroup:2022pth} for the masses of the charged leptons.
\begin{table}[t]
  \centering
  \begin{tabular}{|c|cc|}
  \hline 
  \rule[14pt]{0pt}{0pt}
    Normal Ordering & bfp $\pm1\sigma$ & $3\sigma$ range \\
     \hline
     \rule[14pt]{0pt}{0pt}
    $\sin^2\theta_{12}$ & $0.303^{+0.012}_{-0.012}$ & $0.270 \to 0.341$ \\ \rule[14pt]{0pt}{0pt}
    $\theta_{12}/^\circ$ & $33.41^{+0.75}_{-0.72}$ & $31.31\to35.74$ \\ \rule[14pt]{0pt}{0pt}
    $\sin^2\theta_{23}$ & $0.451^{+0.019}_{-0.016}$ & $0.408\to0.603$ \\ \rule[14pt]{0pt}{0pt}
    $\theta_{23}/^\circ$ & $42.2^{+1.1}_{-0.9}$ & $39.7\to51.0$ \\ \rule[14pt]{0pt}{0pt}
    $\sin^2\theta_{13}$ & $0.02225^{+0.00056}_{-0.00059}$ & $0.02052\to0.02398$ \\ \rule[14pt]{0pt}{0pt}
    $\theta_{13}/^\circ$ & $8.58^{+0.11}_{-0.11}$ & $8.23\to8.91$ \\ \rule[14pt]{0pt}{0pt}
    $\delta_{\mathrm{CP}}/^\circ$ & $232^{+36}_{-26}$ & $144\to350$ \\ \rule[14pt]{0pt}{0pt}
    $\frac{\Delta m_{21}^2}{10^{-5} [\mathrm{eV}^2]}$ & $7.41^{+0.21}_{-0.20}$ & $6.82\to8.03$\\ \rule[16pt]{0pt}{0pt}
    $\frac{\Delta m_{31}^2}{10^{-3} [\mathrm{eV}^2]}$ & $+2.507^{+0.026}_{-0.027}$ & $+2.427\to-2.590$ \rule[-8pt]{0pt}{0pt} \\ \hline 
  \end{tabular}
  \caption{NuFIT 5.2 data~\cite{Esteban:2020cvm,NuFIT5.2}
  in the normal ordering, where $\Delta m^2_{ij}$ is the mass-squared differences between $m_{i}$ and $m_{j}$. }
  \label{tab:NuFIT_data}
\end{table}
% Then, we show the our model parameters.
Our model parameters are
\begin{align}
	v_1, v_2, v_3, y'_{e\mu}, y_{\tau}, y_{\ell1}, y_{\ell2}, \mathrm{X'}_1, m_1, |y'_{De}|, |y'_{D\mu\tau}| \, \mathrm{and} \, \phi_{De}, \notag
\end{align}
where, $v_1$, $v_2$, and $v_3$ are the VEVs of Higgs that satisfy $v_1^2+v_2^2+v_3^2=v^2$ with $v\approx246~\mathrm{GeV}$~\cite{ParticleDataGroup:2022pth}.
For the charged lepton mass matrix $M_{\ell}M_{\ell}^\dagger$, the Yukawa couplings $y_{\ell1}$ and $y_{\ell2}$ are restricted to be between $-\pi$ and $\pi$.
The remaining variables $y'_{e\mu}$, $y_{\tau}$, and $\mathrm{X'}_1$ are determined such that the charged lepton masses $m_e$, $m_\mu$, and $m_\tau$ are reproduced when $M_{\ell}M_{\ell}^\dagger$ is diagonalized.
%*特性方程式を載せても良い
For the neutrino sector, only the two type of mass-squared differences are known, and the absolute masses are not known.
Therefore, of the three neutrino masses $m_1$, $m_2$ and $m_3$, we vary $m_1$ within the range satisfied by the experiment data~\cite{Planck:2018vyg}.
From that the remaining two masses $m_2$ and $m_3$ are fixed.
%m2,m3の式の形を書く

Next, the variables $|y'_{De}|$, $|y'_{D\mu \tau}|$, $\phi_{De}$ of the neutrino mass matrix $m_\nu m_\nu^\dagger$ are determined by reproducing the neutrino masses $m_1, m_2$ and $m_3$ when $m_\nu m_\nu^\dagger$ is diagonalized.
Here, $\phi_{D\mu \tau}$ can be absorbed as a whole phase when calculating $m_\nu m_\nu^\dagger$, so it has no physical meaning.
From the above, we can numerically obtain the unitary matrices $V_\ell$ and $V_\nu$ that diagonalize $M_{\ell}M_{\ell}^\dagger$ and $m_\nu m_\nu^\dagger$.
The PMNS matrix $U$, which represents the lepton flavor mixing is obtained by
\begin{align} \label{eq:PMNS_mat}
	U=V^\dagger_\ell V_\nu.
\end{align}
We denote the $ij$ components of $U$ as $U_{ij}$.
Then we can derive the mixing angles $\theta_{12}$, $\theta_{23}$, and $\theta_{13}$ as follows;
%PDGの3*3のUの表記を書いておく(footnote)
\begin{align}
	\tan \theta_{12} &= \frac{|U_{12}|}{|U_{11}|}, \label{eq:theta_12} \\
	\tan \theta_{23} &= \frac{|U_{23}|}{|U_{33}|}, \label{eq:theta_23} \\
	\sin \theta_{13} &= |U_{13}|. \label{eq:theta_13} 
\end{align}
Therefore, we can get the PMNS matrix $U$ by Eq.~\eqref{eq:PMNS_mat} and also numerically obtain the mixing angles from Eqs.~\eqref{eq:theta_12} - \eqref{eq:theta_13}.
From NuFIT 5.2~\cite{Esteban:2020cvm, NuFIT5.2}, the magnitude of the mixing angles is fixed in the range shown in Table~\ref{tab:NuFIT_data}.

\begin{figure}[t]
%	\begin{center}
%	\includegraphics[width=75mm]{fig_alpha_beta.pdf}
%  	\subcaption{}
%    	\label{fig:alpha_beta}
%	\end{center}
	\begin{minipage}{0.5\hsize}
 		\begin{center}
		\includegraphics[width=70mm]{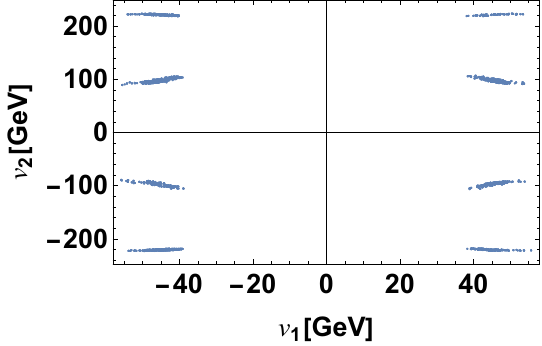}
		\subcaption{}
		\label{fig:v1_v2}
		\end{center}
	\end{minipage}
	\begin{minipage}{0.5\hsize}
		\begin{center}
		\includegraphics[width=70mm]{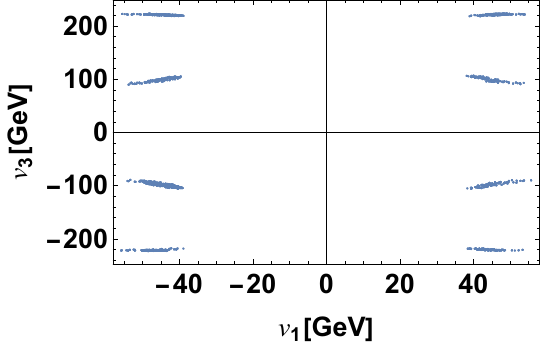}
		\subcaption{}
    		\label{fig:v1_v3}
		\end{center}
	\end{minipage}
	\caption{(a)~The relation between $v_1$ and $v_2$. (b)~The relation between $v_1$ and $v_3$.}
	\label{fig:v}
\end{figure}

Our results that lay within the range of the mixing angles are shown in Table~\ref{tab:NuFIT_data}.
First, the relation of the VEVs of Higgs, $v_1$, $v_2$, and $v_3$ are shown in the Figure~\ref{fig:v}.
The range of $v_1$ is almost $38.18~[\mathrm{GeV}] \leq |v_1| \leq 55.67~[\mathrm{GeV}]$ and takes a relatively small value compared to $v_2$ and $v_3$.
The $v_2$ and $v_3$ are almost $89.51~[\mathrm{GeV}] \sim 106.40~[\mathrm{GeV}]$ or $218.27~[\mathrm{GeV}] \sim 223.30~[\mathrm{GeV}]$.
Our results for the range of $y'_{e\mu}$, $y_{\tau}$, and $\mathrm{X'}_1$ are $6.93\times 10^{-4} \leq y'_{e\mu} \leq 9.91\times 10^{-4}$, $1.99\times 10^{-3}\leq y_{\tau} \leq 1.01 \times 10^{-2}$, and $2.25 \times 10^{-4} \leq \mathrm{X'}_1 \leq 4.21 \times 10^{-2}$ respectively.
Notice the $y'_{e\mu}$ is restricted to a narrow range.
However, $y_{\tau}$ and $\mathrm{X'}_1$ are approximately $10^1$ or $10^2$ wide.
Then, the results for the ranges $|y'_{De}|$ and $|y'_{D\mu\tau}|$ are $7.90 \times 10^{-13}~[\mathrm{eV}^{-1/2}] \leq |y'_{De}| \leq 8.80 \times 10^{-13}~[\mathrm{eV}^{-1/2}]$ and $1.81 \times 10^{-12}~[\mathrm{eV}^{-1/2}] \leq |y'_{D\mu\tau}| \leq 1.84 \times 10^{-12}~[\mathrm{eV}^{-1/2}] $ respectively.
If we consider the Yukawa couplings of Dirac neutrinos are on the order of $10^{-1}$, the masses of the right-handed Majorana neutrinos, $M_{eR}$ and $M_{\mu\tau R}$, are on the order of $10^{15}~[\mathrm{GeV}]$ and $10^{13}~[\mathrm{GeV}]$.
Recall we are taking the full range of $-\pi$ to $\pi$ for $y_{\ell1}$ and $y_{\ell2}$.
The results of $m_1$ and $\phi_{De}$ will be shown later. 

\begin{figure}[t]
	\begin{center}
	\includegraphics[width=120mm]{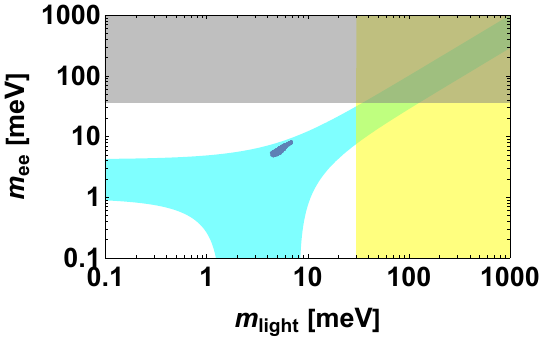}
	\end{center}
	\caption{The relation between the lightest neutrino mass $m_{light}$ and the effective Majorana neutrino mass $m_{ee}$ at the $0\nu\beta\beta$ decay experiment. The blue area is model independent analyzes for the in normal ordering of the neutrino mass hierarchy. The gray area is upper limit on the effective Majorana neutrino mass of $36~[\mathrm{meV}]$ at $90\%$ C.L.~\cite{KamLAND-Zen:2022tow}. The yellow area is upper limit on the lightest neutrino mass of $30.1~[\mathrm{meV}]$ which is estimated in~\cite{Planck:2018vyg}.}
	\label{fig:mlight_mee}
\end{figure}

\begin{figure}[t]
	\begin{minipage}{0.5\hsize}
 		\begin{center}
		\includegraphics[width=75mm]{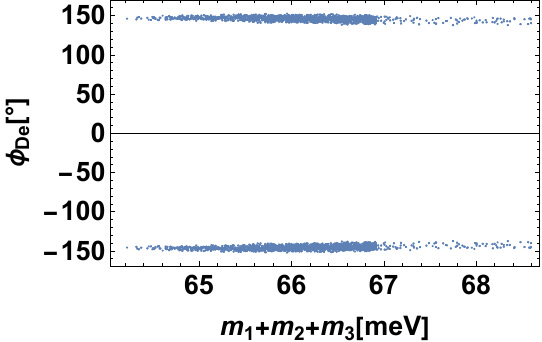}
		\subcaption{}
		\label{fig:summn_phiDe}
		\end{center}
	\end{minipage}
	\begin{minipage}{0.5\hsize}
		\begin{center}
		\includegraphics[width=80mm]{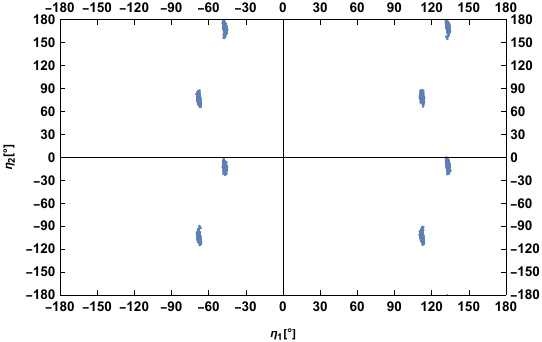}
		\subcaption{}
    		\label{fig:eta1_eta2}
		\end{center}
	\end{minipage}
	\caption{(a)~The relation between the sum of neutrino masses and the complex phase of the Yukawa coupling of Dirac neutrino $\phi_{De}$. (b)~The relation between Majorana phases $\eta_1$ and $\eta_2$.}
\end{figure}

Next, within the allowed parameter region from above, we predict the lepton mixing angles, the Dirac CP phase $\delta_{CP}$, the effective Majorana neutrino mass $m_{ee}$ for $0\nu\beta\beta$ decay experiments, the sum of neutrino masses, and the Majorana phases $\eta_1, \eta_2$.
\begin{figure}[t]
	\begin{center}
	\includegraphics[width=120mm]{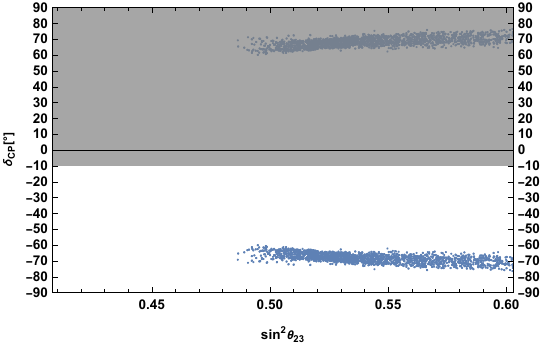}
	\end{center}
	\caption{The relation between $\sin^2 \theta_{23}$ and $\delta_{CP}$. In our model, the positive and negative CP phases are obtained equally. The white area is the 3$\sigma$ range from NuFIT 5.2~\cite{Esteban:2020cvm,NuFIT5.2}.}
	\label{fig:sin23sq_deltaCP}
\end{figure}
Our prediction for the mixing angle $\sin^2 \theta_{23}$ and Dirac CP phase $\delta_{CP}$ is shown in the Figure~\ref{fig:sin23sq_deltaCP}.
The Dirac CP phase $\delta_{CP}$ can be obtained from the two equations.
One equation is the Jarlskog invariant~\cite{Jarlskog:1985ht} $J_{CP}$, which is written as
\begin{align}
	J_{CP}=\mathrm{Im}[U_{11}U^*_{21}U^*_{12}U_{22}].
\end{align}
Then, we can get $\delta_{CP}$ by using $J_{CP}$,
\begin{align} \label{eq:sin_delta_CP}
	\sin \delta_{CP} = \frac{J_{CP}}{s_{12} c_{12} s_{23} c_{23} s_{13} c^2_{13} },
\end{align}
where, $s_{ij}=\sin \theta_{ij}$ and $c_{ij}=\cos \theta_{ij}$.
The other equation to obtain $\delta_{CP}$ is
\begin{align} \label{eq:cos_delta_CP}
	\cos \delta_{CP} = \frac{s^2_{12} s^2_{23} + c^2_{12} s^2_{13} c^2_{23} - |U_{31}|^2 }{2s_{12} c_{12} s_{23} c_{23} s_{13} }.
\end{align}
% From Eq.~\eqref{eq:sin_delta_CP} and Eq.~\eqref{eq:cos_delta_CP}, we obtain $\delta_{CP}$.
We find the $|\delta_{CP}|$ in our model falls in the range $60.12^\circ \sim 76.47^\circ$.
In addition, we can predict $\sin\theta_{23}$ to be in the range $0.486 \sim 0.603$.
The results of $\sin\theta_{12}$ and $\sin\theta_{13}$ take the full range of possible values.
Then, the prediction of the lightest neutrino mass $m_{light}$ and the effective Majorana neutrino mass $m_{ee}$ at the $0\nu\beta\beta$ decay experiment is shown in the Figure~\ref{fig:mlight_mee}.
We derive the effective Majorana neutrino mass $m_{ee}$ using 
\begin{align}
	m_{ee} = |\Sigma_i U^2_{1i} m_i|.
\end{align}
We find $m_{ee}\simeq6.33[\mathrm{meV}]$ and $m_{light}\simeq5.53[\mathrm{meV}]$.
Both of those results lay relatively close to upper limits.
Therefore, we consider this model is a relatively easy to confirm by the near future experiments.
Lastly, the relation between the sum of neutrino masses and the complex phase of the Yukawa coupling of Dirac neutrino $\phi_{De}$ is shown in the Figure~\ref{fig:summn_phiDe}.
For the comparison of the sum of neutrino masses, we find the sum of neutrino masses $m_1+m_2+m_3\simeq66.1~[\mathrm{meV}]$, which is much smaller than the experimental limit from cosmology~\cite{Planck:2018vyg} is $0.12~[\mathrm{eV}]$.
% Therefore, the sum of neutrino masses predicted by our model is much smaller than the experimental limit. 
We derived $|\phi_{De}|$ to be approximately $146^\circ$.
Finally, we show the prediction of Majorana phases $\eta_1$ and $\eta_2$ in the Figure~\ref{fig:eta1_eta2}.
The Majorana phases $\eta_1$ and $\eta_2$ can be derived as
\begin{align}
	e^{i\eta_1}=\frac{U_{11}U^*_{13}}{c_{12}c_{13}s_{13}e^{i\delta_{CP}}}, \quad e^{i\eta_2}=\frac{U_{12}U^*_{13}}{s_{12}c_{13}s_{13}e^{i\delta_{CP}}}. 
\end{align}
Because it has not yet been determined whether the neutrino is a Dirac or Majorana particle, this is only a possible prediction for magnitude of the Majorana phases $\eta_1$ and $\eta_2$.

%------------------------------------------------------------------------------%
\section{Potential analysis}
\label{sec:poten_analysis}
%------------------------------------------------------------------------------%
In this section, we analyze the scalar potential.
The scalar potential in our model (Table~\ref{tab:model}) can be written as,
\begin{align} \label{eq:poten_Higgs_revised}
	V = \mu^2 \phi^\dagger \phi + \lambda (\phi^\dagger \phi)^2 + c\phi^\dagger \phi X^\dagger X + k\phi^\dagger \phi \Theta^\dagger \Theta + (g\phi^\dagger \phi \Theta^\dagger X + h.c.),
\end{align}
where $\lambda>0$.
We calculate this potential using the multiplication rules of the $S_4$ symmetry in Eq.~\eqref{eq:multi_rule_rep}-\eqref{eq:multi_rule_3*3} from Appendix \ref{sec:intro_S4}.
Specifically, from Eq.~\eqref{eq:poten_cal_3HDM_with_S4} in Appendix \ref{sec:3HDM_S4}, the first and second term in Eq.~\eqref{eq:poten_Higgs_revised} are derived as below,
\begin{align}
	\mu^2 \phi^\dagger \phi + \lambda (\phi^\dagger \phi)^2 =& \mu^2(|\phi_1|^2+|\phi_2|^2+|\phi_3|^2)\nonumber \\
	&+(\lambda_1+\frac{2}{3}\lambda_2)(|\phi_1|^4+|\phi_2|^4+|\phi_3|^4)\notag \\
	&+(2\lambda_1-\frac{2}{3}\lambda_2+2\lambda_3-2\lambda_4)(|\phi_1\phi_2|^2+|\phi_2\phi_3|^2+|\phi_3\phi_1|^2) \notag \\
	&+(\lambda_3+\lambda_4)\left[(\phi_1\phi_2^\dag)^2+(\phi_2\phi_3^\dag)^2+(\phi_3\phi_1^\dag)^2+h.c.\right].
\end{align}
Then, we obtain the third term in Eq.~\eqref{eq:poten_Higgs_revised},
\begin{align}
	c\phi^\dagger \phi X^\dagger X =& c \begin{pmatrix} \phi^\dagger_1 \\ \phi^\dagger
_2 \\ \phi^\dagger_3 \end{pmatrix} \times \begin{pmatrix} \phi_1 \\ \phi
_2 \\ \phi_3 \end{pmatrix} \times \begin{pmatrix} X^\dagger_1 \\ X^\dagger_2 \end{pmatrix} \times \begin{pmatrix} X_1 \\ X_2 \end{pmatrix} \notag \\
	=& c_1(|\phi_1|^2+|\phi_2|^2+|\phi_3|^2)(|X_1|^2+|X_2|^2) \notag \\
	&+ c_2\left[\frac{1}{\sqrt{2}} (|\phi_2|^2-|\phi_3|^2)(X_1^\dagger X_2 + X^\dagger_2 X_1)+\frac{1}{\sqrt{6}}(2|\phi_1|^2+|\phi_2|^2+|\phi_3|^2)(|X_1|^2-|X_2|^2)\right].
\end{align}
Similarly, we obtain the forth term and fifth term in Eq.~\eqref{eq:poten_Higgs_revised},
\begin{align}
	k\phi^\dagger \phi \Theta^\dagger \Theta =& k \begin{pmatrix} \phi^\dagger_1 \\ \phi^\dagger
_2 \\ \phi^\dagger_3 \end{pmatrix} \times \begin{pmatrix} \phi_1 \\ \phi
_2 \\ \phi_3 \end{pmatrix} |\Theta|^2 \notag \\
	=& k|\Theta|^2 (|\phi_1|^2+|\phi_2|^2+|\phi_3|^2),
\end{align}
\begin{align}
	g(\phi^\dagger \phi \Theta^\dagger X + h.c.)=& \left(g \begin{pmatrix} \phi^\dagger_1 \\ \phi^\dagger
_2 \\ \phi^\dagger_3 \end{pmatrix} \times \begin{pmatrix} \phi_1 \\ \phi
_2 \\ \phi_3 \end{pmatrix} \Theta^\dagger \begin{pmatrix} X_1 \\ X_2 \end{pmatrix} +h.c.\right) \notag \\
	=& \left[ g \left(\frac{1}{\sqrt{2}}(|\phi_2|^2-|\phi_3|^2) \Theta^\dagger X_1 +\frac{1}{\sqrt{6}}(2|\phi_1|^2+|\phi_2|^2+|\phi_3|^2) \Theta^\dagger X_2 \right) + h.c.\right].
\end{align}
Here, we assume the VEVs of $X$ and $\Theta$ to $(\mathrm{X}_1,0)$ and $\Theta_0$, and the VEV of $\phi_i$ is $(0, v_i/\sqrt{2}) \, (i=1,2,3)$.
Then, we consider the minimum conditions of the potential, and we obtain the following equations,
\begin{align}
	v_1 &= \pm\sqrt{-\frac{\mu^2+c_1 \mathrm{X}_1^2+ k\Theta_0^2 }{3\lambda_1 + 4\lambda_3} -\frac{2}{3\sqrt{6}} \frac{3\lambda_1+2\lambda_2}{(3\lambda_1+4\lambda_4)(\lambda_2-2\lambda_3)}c_2 \mathrm{X}_1^2 },  \\
	v_2 &= \pm\sqrt{-\frac{\mu^2+c_1 \mathrm{X}_1^2+ k\Theta_0^2 }{3\lambda_1 + 4\lambda_3} +\frac{1}{3\sqrt{6}} \frac{3\lambda_1-4\lambda_2+12\lambda_3}{(3\lambda_1+4\lambda_4)(\lambda_2-2\lambda_3)}c_2 \mathrm{X}_1^2 - \frac{1}{\sqrt{2}} \frac{g\Theta_0 \mathrm{X}_1}{\lambda_2-2\lambda_3} },  \\
	v_3 &= \pm\sqrt{-\frac{\mu^2+c_1 \mathrm{X}_1^2+ k\Theta_0^2 }{3\lambda_1 + 4\lambda_3} +\frac{1}{3\sqrt{6}} \frac{3\lambda_1-4\lambda_2+12\lambda_3}{(3\lambda_1+4\lambda_4)(\lambda_2-2\lambda_3)}c_2 \mathrm{X}_1^2 + \frac{1}{\sqrt{2}} \frac{g\Theta_0 \mathrm{X}_1}{\lambda_2-2\lambda_3} }. 
\end{align}
Since there are many parameters for the scalar potential, the result in the Figure~\ref{fig:v} can be realized.
One worry could be how the masses of the neutral and charged Higgs contribute to the flavor-changing neutral currents(FCNC).
We expect the masses can be heavy enough to suppress the FCNC and use this model by adjusting the parameters of the potential.
However, that is beyond the scope of this work.

\section{Summary}
\label{sec:Summary}
We have built a lepton flavor model with $S_4$ and $U(1)_{\mathrm{FN}}$ symmetries.
The left-handed leptons are assigned as an $S_4$ triplet and the right-handed tauon as an $S_4$ singlet.
Then, we assign the right-handed electron and muon as an $S_4$ doublet and have $U(1)_{\mathrm{FN}}$ charge.
We have assumed there are three right-handed Majorana neutrinos that have an $S_4$ symmetric charge and two types of scalar fields to reproduce the lepton flavor mixing.
We have introduced three Higgs doublets that were assigned as an $S_4$ triplet, and calculated the charged lepton mass matrix and the left-handed Majorana neutrino mass matrix.

We have performed a numerical analysis to obtain the PMNS matrix, and gave predictions for the mixing angle $\theta_{23}$ and the Dirac CP phase $\delta_{CP}$.
The results were $0.486<\sin\theta_{23}<0.603$ and $60.12^\circ<|\delta_{CP}|<76.47^\circ$, which is a strong prediction for the $|\delta_{CP}|$.
We also gave predictions for the lightest neutrino mass $m_{light}$, the effective Majorana neutrino mass $m_{ee}$ at the $0\nu\beta\beta$ decay experiment, the sum of neutrino masses and two Majorana phases $\eta_1$ and $\eta_2$.
The results were $m_{light}\simeq5.53~[\mathrm{meV}]$, $m_{ee}\simeq6.33~[\mathrm{meV}]$, $m_1+m_2+m_3\simeq66.1~[\mathrm{meV}]$.
For $m_{ee}$ and $m_{light}$, we were able to obtain values relatively close to current experimentally upper limits.
Finally, we analyzed the scalar potential to get the conditions for the Higgs VEVs.
% We got the conditions about the Higgs VEVs.

%Thus, we have built a flavor model with $S_4$ and $U(1)_{\mathrm{FN}}$ symmetries and 3HDM in the lepton sector.
Our model could be extended to the quark sector.
% We can extend the model to the quark sector.
Where we could impose the $S_4$ symmetry on the quarks as well, and consider how to assign the $S_4$ charge and built a flavor model.
The model would be tested by the precise observations of the CP phase in the quark sector.
It is interesting to study more phenomenological aspects on the flavor physics based on flavor symmetry with multi-Higgs in the near future.

%-------- acknowledgement -------%
\vspace{1cm}
\noindent
{\large \bf Acknowledgement}
\vspace{1mm}

We thank M. Tanimoto, Y. Shimizu, N. Benoit, Y. Watanabe and S. Takeshita for useful discussions.
This work was supported by JST, the establishment of university fellowships towards the creation of science technology innovation, Grant Number JPMJFS2129.

%\newpage 
%---------------------------------------------------------%
%--------------- Appendix --------------------------------%
%---------------------------------------------------------%
\appendix
\section*{Appendix}
%---------------------------------------------------------%
%--------------- S4 sym. -----------------------------------%
%---------------------------------------------------------%
\section{$S_4$ symmetry}
\label{sec:intro_S4}
We give a brief introduction of the $S_4$ symmetry.
%S4の論文を載せる
The $S_4$ symmetry is the symmetry of the $S_4$ group, which is the symmetric group of order 4.
The $S_4$ group consists of all permutations among four objects $(x_1, x_2, x_3, x_4)$.
Therefore, the number of elements of the $S_4$ group are $4!=24$
\begin{align}
	(x_1, x_2, x_3, x_4) \rightarrow (x_i, x_j, x_k, x_l).
\end{align}
The $S_4$ group is the smallest non-Abelian discrete symmetry group that has five irreducible representations.
There are two types of singlets $\mathbf{1}$, $\mathbf{1'}$, one type of doublet $\mathbf{2}$, and two types of triplets $\mathbf{3}$, $\mathbf{3'}$.
The $S_4$ symmetry represents the symmetry of cubic geometry.

Next, we show the multiplication rule of the $S_4$ symmetry:
\begin{align} \label{eq:multi_rule_rep}
	\mathbf{3} \otimes \mathbf{3} &= \mathbf{1} \oplus \mathbf{2} \oplus \mathbf{3} \oplus \mathbf{3'}, \notag \\
	\mathbf{3'} \otimes \mathbf{3'} &= \mathbf{1} \oplus \mathbf{2} \oplus \mathbf{3} \oplus \mathbf{3'}, \notag \\
	\mathbf{3} \otimes \mathbf{3'} &= \mathbf{1'} \oplus \mathbf{2} \oplus \mathbf{3} \oplus \mathbf{3'}, \notag \\
	\mathbf{2} \otimes \mathbf{2} &= \mathbf{1} \oplus \mathbf{1'} \oplus \mathbf{2}, \notag \\
	\mathbf{2} \otimes \mathbf{3} &= \mathbf{3} \oplus \mathbf{3'}, \\
	\mathbf{2} \otimes \mathbf{3'} &= \mathbf{3} \oplus \mathbf{3'}, \notag \\
	\mathbf{3} \otimes \mathbf{1'} &= \mathbf{3'},  \notag \\
	\mathbf{3'} \otimes \mathbf{1'} &= \mathbf{3}, \notag \\
	\mathbf{2} \otimes \mathbf{1'} &= \mathbf{2}, \notag
\end{align}
\begin{align} \label{eq:multi_rule_2*2}
	\begin{pmatrix}
		a_1\\
		a_2
	\end{pmatrix}_\mathbf{2}
	\otimes
	\begin{pmatrix}
		b_1\\
		b_2
	\end{pmatrix}_\mathbf{2}
	&=\left (a_1b_1+a_2b_2\right )_\mathbf{1}
	\oplus
	\left (-a_1b_2+a_2b_1\right )_\mathbf{1'} \notag \\
	&\oplus
	\begin{pmatrix}
		a_1b_2+a_2b_1 \\
		a_1b_1-a_2b_2
	\end{pmatrix}_\mathbf{2}, 
\end{align}
\begin{align} \label{eq:multi_rule_3*3}
	\begin{pmatrix}
		a_1\\
		a_2\\
		a_3
	\end{pmatrix}_\mathbf{3}
	\otimes
	\begin{pmatrix}
		b_1\\
		b_2\\
		b_3
	\end{pmatrix}_\mathbf{3}
	&=\left (a_1b_1+a_2b_2+a_3b_3\right )_\mathbf{1} \notag \\
	&\oplus
	\begin{pmatrix}
		\frac{1}{\sqrt{2}}(a_2b_2-a_3b_3) \\
		\frac{1}{\sqrt{6}}(-2a_1b_1+a_2b_2+a_3b_3)
	\end{pmatrix}_\mathbf{2} \notag \\
	&\oplus
	\begin{pmatrix}
		a_3b_2+a_2b_3 \\
		a_1b_3+a_3b_1 \\
		a_2b_1+a_1b_2
	\end{pmatrix}_\mathbf{3}
	\oplus
	\begin{pmatrix}
		a_3b_2-a_2b_3 \\
		a_1b_3-a_3b_1 \\
		a_2b_1-a_1b_2
	\end{pmatrix}_\mathbf{3'}.
\end{align}
More details are shown in the review~\cite{Ishimori:2010au,Ishimori:2012zz}.

%----------------------------------------------------------------------%
%----------------------    3HDM_S4    -----------------------------------%
%----------------------------------------------------------------------%
\section{3HDM with $S_4$ symmetry} \label{sec:3HDM_S4}
We briefly introduce the 3HDM with $S_4$ symmetry.
In general 3HDM means that the SM Higgs sector is extended to include three Higgs doublets, denoted as
\begin{align}
	\phi_1=\begin{pmatrix} \phi^+_1 \\ \phi^0_1 \end{pmatrix}, \phi_2=\begin{pmatrix} \phi^+_2 \\ \phi^0_2 \end{pmatrix}, \phi_3=\begin{pmatrix} \phi^+_3 \\ \phi^0_3, \end{pmatrix} 
\end{align}
where the Higgs doublets $\phi_i$, with $i = 1$, $2$, $3$, have the same $U(1)_Y$ hypercharge.
One Higgs doublet consists of four real scalar fields.
Thus, there are twelve real scalar fields in the 3HDM.

In this study, we consider the Higgs fields to also have a flavor symmetry.
Therefore, we regard three Higgs fields as a $S_4$ triplet,
\begin{align}
	\phi=\begin{pmatrix} \phi_1 \\ \phi_2 \\ \phi_3 \end{pmatrix}.
\end{align}
In general the scalar potential is given by
\begin{align}
	V = \mu^2 \phi^\dagger \phi + \lambda (\phi^\dagger \phi)^2, 
\end{align}
where, $\mu^2<0, \lambda>0$.
We get this potential by using the multiplication rule of the $S4$ symmetry
\begin{align}
	V=&\mu^2
	\begin{pmatrix}
	\phi^\dagger_1 \\
	\phi^\dagger_2 \\
	\phi^\dagger_3
	\end{pmatrix}
	\times
	\begin{pmatrix}
	\phi_1 \\
	\phi_2 \\
	\phi_3
	\end{pmatrix}
	+\lambda
	\begin{pmatrix}
	\phi^\dagger_1 \\
	\phi^\dagger_2 \\
	\phi^\dagger_3
	\end{pmatrix}
	\times
	\begin{pmatrix}
	\phi_1 \\
	\phi_2 \\
	\phi_3
	\end{pmatrix}
	\times
	\begin{pmatrix}
	\phi^\dagger_1 \\
	\phi^\dagger_2 \\
	\phi^\dagger_3
	\end{pmatrix}
	\times
	\begin{pmatrix}
	\phi_1 \\
	\phi_2 \\
	\phi_3
	\end{pmatrix} \notag \\
	=&\mu^2 (|\phi_1|^2+|\phi_2|^2+|\phi_3|^2) + \lambda_1 (|\phi_1|^2+|\phi_2|^2+|\phi_3|^2)^2 \notag \\
	&+\frac{2}{3} \lambda_2 (|\phi_1|^4+|\phi_2|^4+|\phi_3|^4-|\phi_1|^2|\phi_2|^2-|\phi_2|^2|\phi_3|^2-|\phi_3|^2|\phi_1|^2) \notag \\
	&+\lambda_3 \left[(\phi^\dagger_1 \phi_2)^2 + (\phi^\dagger_2 \phi_3)^2 + (\phi^\dagger_3 \phi_1)^2 + |\phi_1|^2|\phi_2|^2 + |\phi_2|^2|\phi_3|^2 + |\phi_3|^2|\phi_1|^2 + h.c.\right] \notag \\
	&+\lambda_4 \left[(\phi^\dagger_1 \phi_2)^2 + (\phi^\dagger_2 \phi_3)^2 + (\phi^\dagger_3 \phi_1)^2 - |\phi_1|^2|\phi_2|^2 - |\phi_2|^2|\phi_3|^2 - |\phi_3|^2|\phi_1|^2 + h.c.\right] \notag \\
	=&\mu^2(|\phi_1|^2+|\phi_2|^2+|\phi_3|^2)\nonumber \\
	&+(\lambda_1+\frac{2}{3}\lambda_2)(|\phi_1|^4+|\phi_2|^4+|\phi_3|^4)\notag \\
	&+(2\lambda_1-\frac{2}{3}\lambda_2+2\lambda_3-2\lambda_4)(|\phi_1\phi_2|^2+|\phi_2\phi_3|^2+|\phi_3\phi_1|^2) \notag \\
	&+(\lambda_3+\lambda_4)\left[(\phi_1\phi_2^\dag)^2+(\phi_2\phi_3^\dag)^2+(\phi_3\phi_1^\dag)^2+h.c.\right]. \label{eq:poten_cal_3HDM_with_S4}
\end{align}
Then, we may write the VEVs of Higgs doublets $\phi_i$ where $i = 1$, $2$, $3$ as
\begin{align} \label{eq:VEV_Higgs}
	 \langle \phi_i \rangle = \begin{pmatrix} 0 \\ \frac{1}{\sqrt{2}}v_i \end{pmatrix}. 
\end{align} 
%\begin{align}
%	\phi_i=\begin{pmatrix} \phi^+_i \\ \frac{1}{\sqrt{2}}(v_i + \rho +i \sigma) \end{pmatrix} 
%\end{align}
By taking into account the minimization conditions on the potential in Eq.~\eqref{eq:poten_cal_3HDM_with_S4}, the following relation about the VEVs of Higgs is obtained,
\begin{align}
	v_1^2=v_2^2=v_3^2=\frac{-\mu^2}{3\lambda_1+4\lambda_3}.
\end{align}
We have considered the VEVs to be real numbers.

%----------------------------------------------------------------------%
%----------------------    References    ----------------------------------%
%----------------------------------------------------------------------%
%\newpage

\end{document}